\documentclass{DISproc}

\usepackage{relsize}
\usepackage{amsmath}
\usepackage{wrapfig}
\usepackage{amssymb}
\usepackage{lineno}
\def\babar{\mbox{\slshape B\kern-0.1em{\smaller A}\kern-0.1em B\kern-0.1em{\smaller A\kern-0.2em R }}} %

\begin{document}

%------------------------------------
\title{Search for low-mass Higgs states @ \babar}

%for single authors the superscripts are optional
\author{{\slshape Valentina Santoro$^1$}\\[1ex]
$^1$INFN Ferrara, via Saragat~1, 44122 Ferrara, Italy\\ }

% please enter the contribution ID for the DOI
\contribID{xy}

\doi  % if there is an online version we will register DOIs

\maketitle

\begin{abstract}
Several types of new-physics models predict the existence of light dark matter candidates and low-mass Higgs states. Previous \babar searches for  invisible light-Higgs decays have excluded large regions of model parameter space. We present searches for a dark-sector Higgs produced in association with a dark gauge boson and searches for a light Higgs in $\Upsilon (nS)$ decays.
\end{abstract}

\section{Search for Low-Mass Dark Matter at \babar}

\begin{figure}[htb]
  \centering
  \includegraphics[width=0.5\textwidth]{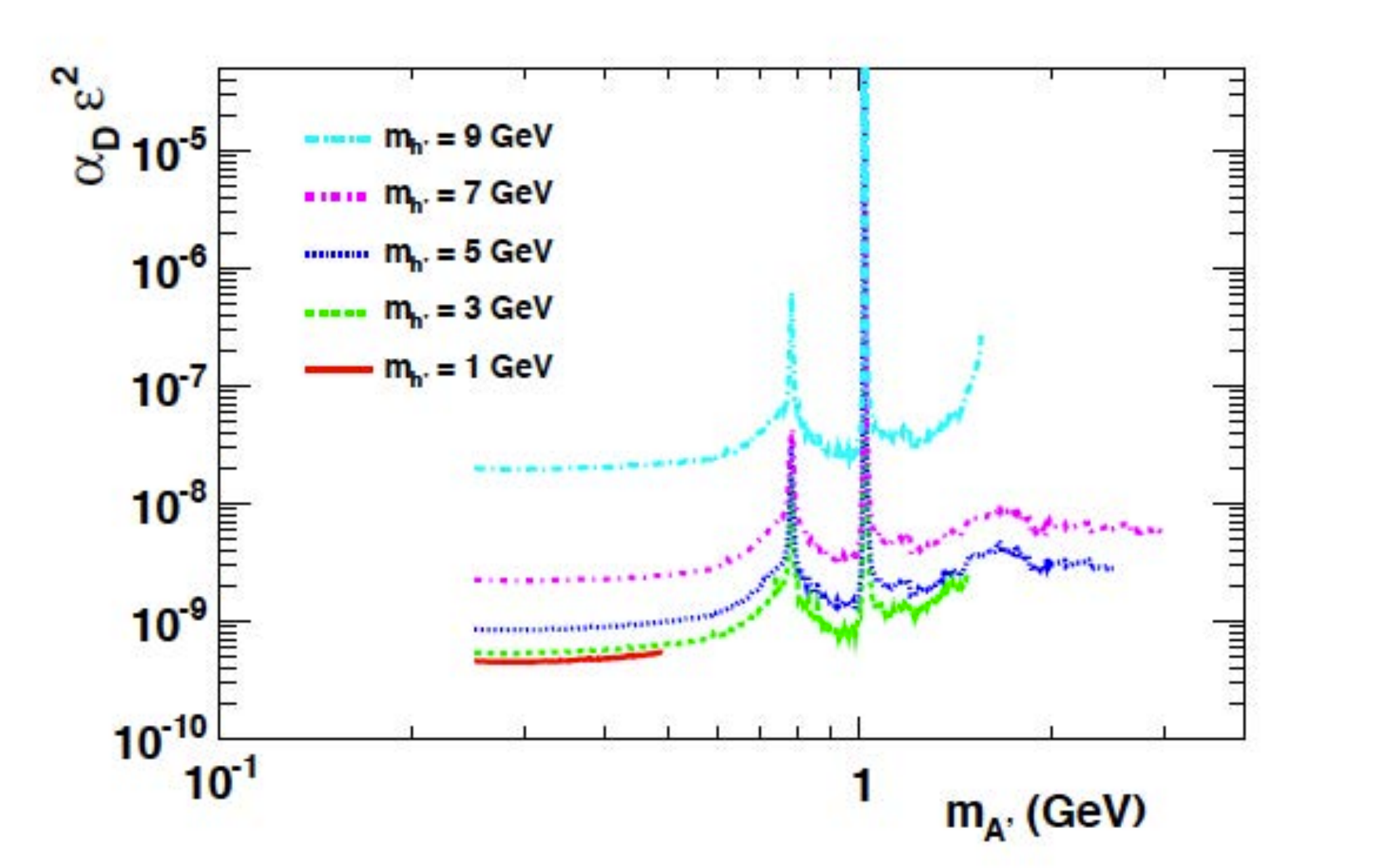}
  \includegraphics[width=0.47\textwidth]{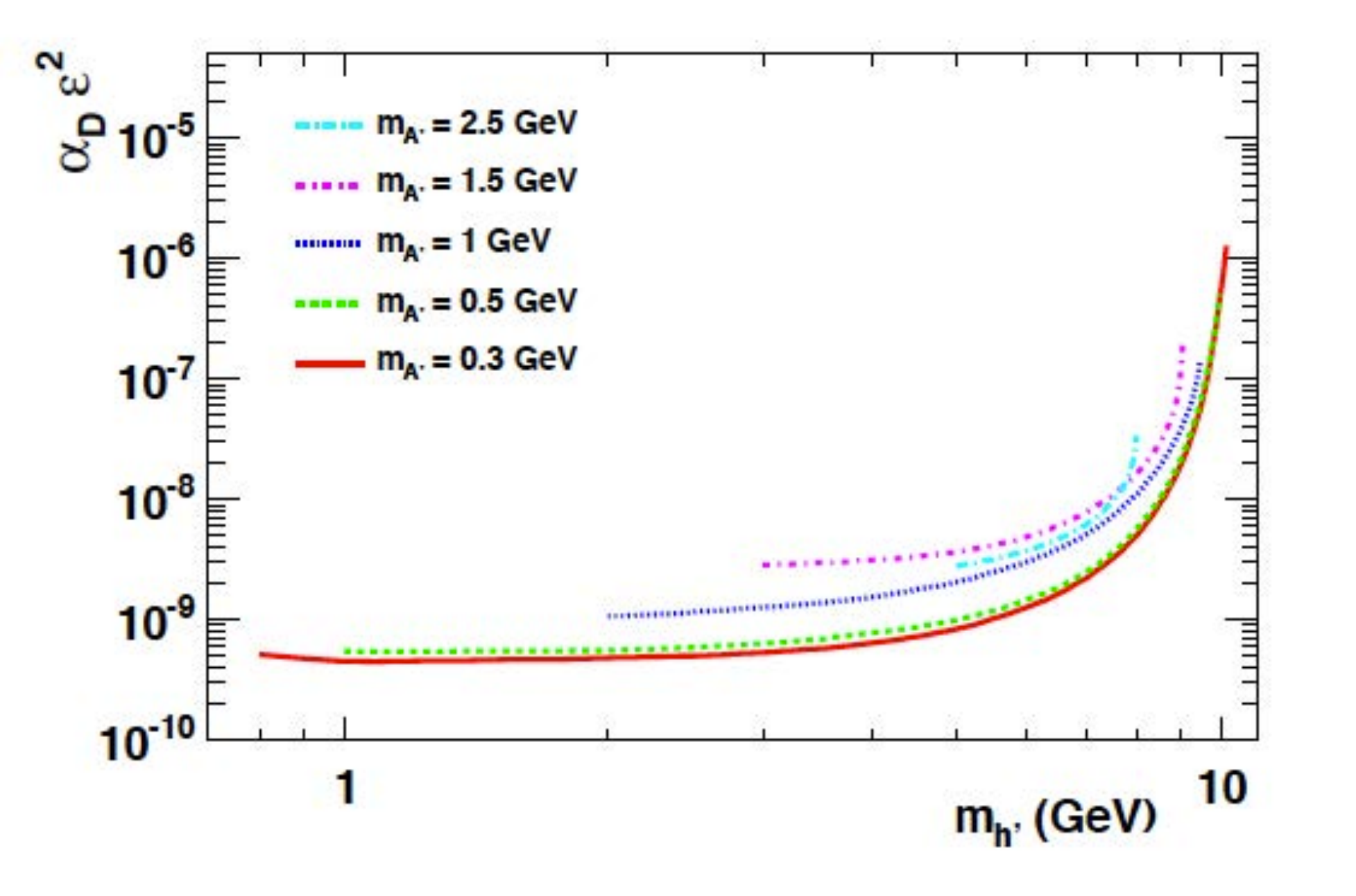}
  \caption{Limits on $\alpha_{\mathrm D} \epsilon^{2}$ for the Dark Photon (left) and Dark Higgs (right).}
  \label{Fig:1}
\end{figure}

\begin{figure}[htb]
  \centering
  \includegraphics[width=0.5\textwidth]{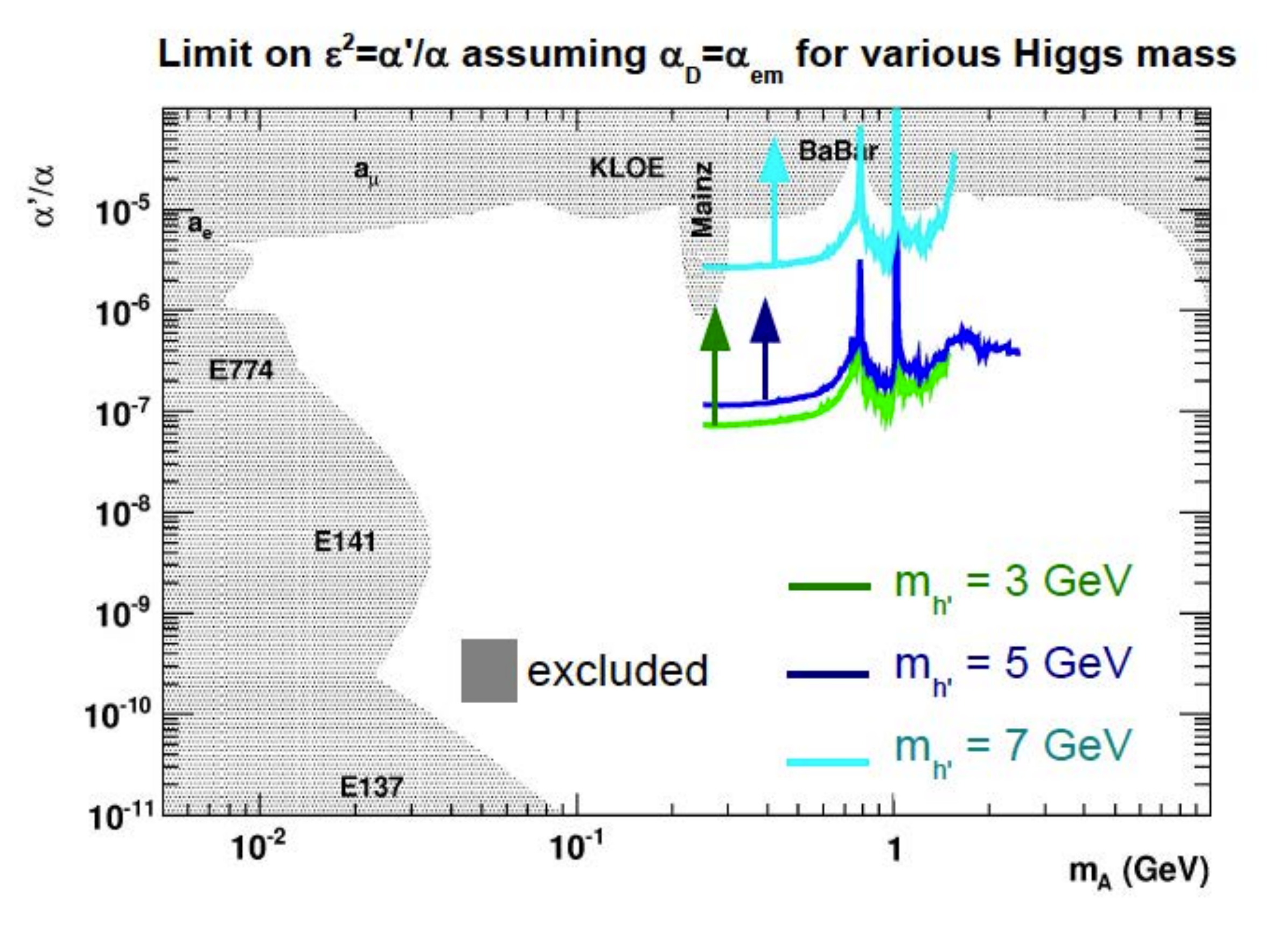}
   \caption{Limit on the mixing strength, $\epsilon^{2}$, for various Higgs masses.}
  \label{Fig:2}
\end{figure}

We have now an  overwhelming astrophysical evidence of dark matter. To explain this experimental evidence many theoretical models \cite{Batell:2009yf}  introducing new dark forces mediated by new gauge boson have been introduced.  These models propose Weakly Interacting Massive Particles (WIMPs) that can annihilate into pairs of dark bosons, which subsequently annihilate to lepton pairs (protons are kinematically forbidden).
One of this model \cite{ArkaniHamed:2008qn,Essig:2009nc}  introduce a new dark sector that couples to the SM with a dark boson (i.e. the dark photon $A{'}$) through a small kinetic mixing term. Thanks to their large luminosities and low-background environment the $B$-factories offer an ideal place to probe for MeV-GeV dark matter, complementing searches from LHC.
The dark boson mass is generated via the Higgs mechanism, adding a dark Higgs boson ($h^{'}$) to the theory. The dark photon and the Higgs bosons could have a comparable mass (GeV-scale). A very minimal scenario has a single dark photon and a single dark Higgs boson.\\
 In the \babar analysis \cite{bba} we make use of the $Higgsstrahlung$ process $e^{+}e^{-}\to A^{'*} \to A^{'}h^{'},~h^{'}\to A^{'}A^{'}$ using 521 $fb^{-1}$. The signal is either fully reconstructed into lepton or pion pair (exclusively mode), or partially reconstructed (inclusive mode). Only two of the three hidden photons are reconstructed in the latter case, and the four-momentum of the third one is identified to that of the recoiling system.\\
In these searches no significant signal is observed. Using uniform priors in the cross section upper limits on the  $e^{+}e^{-}\to A^{'*} \to A^{'}h^{'},~h^{'}\to A^{'}A^{'}$ cross section are obtained as a function of the hidden Higgs and hidden photon masses. These limits on the cross section are translated into 90 \% upper limit on $\alpha_{\mathrm{D}}\epsilon^{2}$  shown in Fig.~\ref{Fig:1}, where $\alpha_{\mathrm{D}}=g_{\mathrm{D}}2/4\pi$, $g_{\mathrm{D}}$ is the dark sector gauge coupling, $\epsilon$ is the mixing strength. Values down to $10^{-10}$ - $10^{-8}$ are excluded for a large range of hidden photon and hidden Higgs masses, assuming prompt decay. Assuming   $\alpha_{\mathrm{D}}=\alpha \simeq 1/137$, limits on the mixing strength in the range $10^{-4}$ - $10^{-3}$ are derived as shown in  Fig.~\ref{Fig:2}, these limits are an order of magnitude smaller than the current experimental bounds extracted from direct photon production in this mass range.

\section{Search for light Higgs @ \babar}
\begin{figure}[htb]
  \centering
  \includegraphics[width=0.47\textwidth]{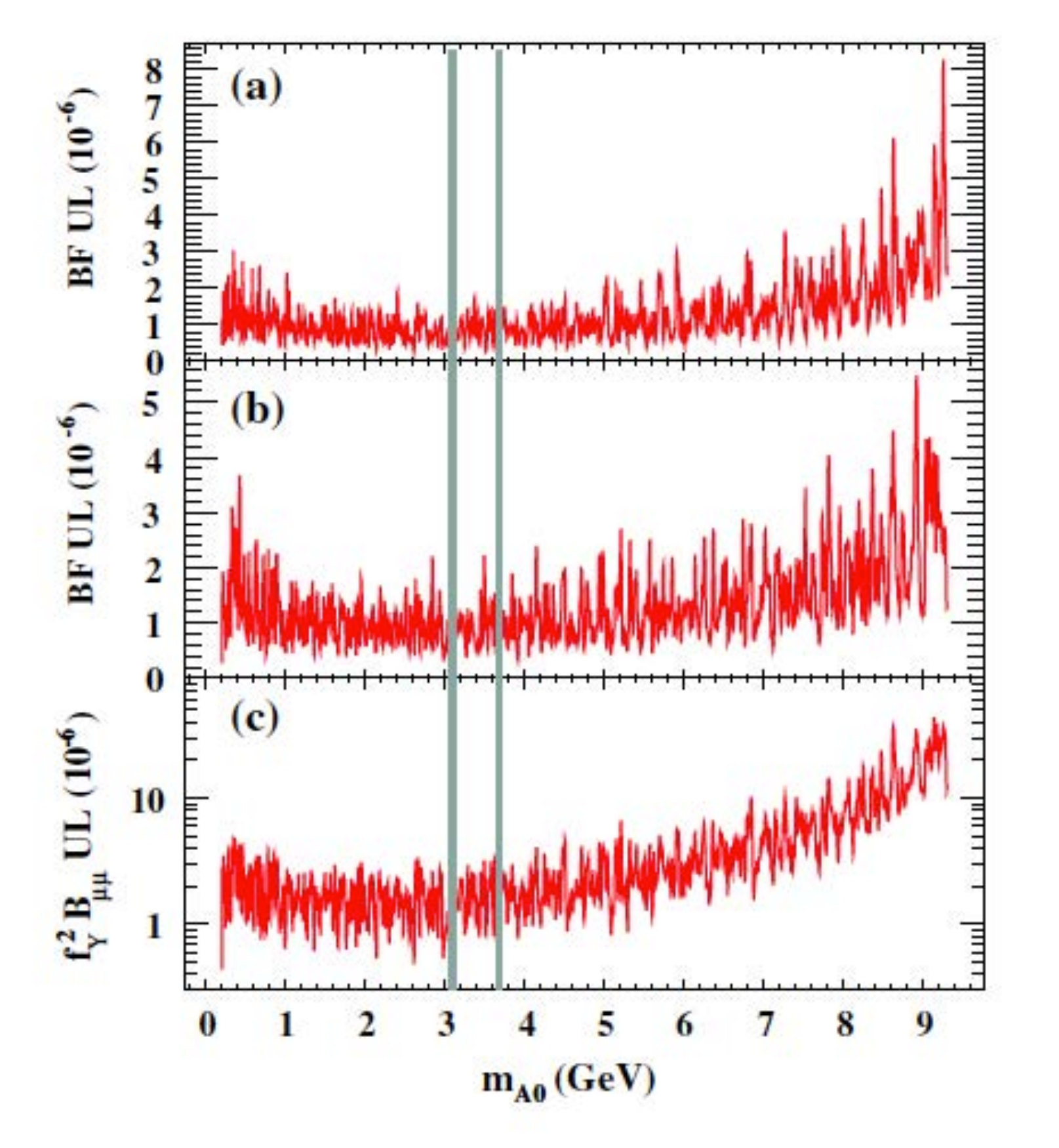}
  \caption{90 \% C.L. upper limits on (a) $ {\cal B}  (\Upsilon(2S)\to \gamma A^{0}) \times \cal{B}_{\mu^{+}\mu^{-}}$, (b) $ {\cal B}  (\Upsilon(3S)\to \gamma A^{0}) \times \cal{B}_{\mu^{+}\mu^{-}}$} and (c) effective coupling $f^{2}_{\Upsilon} \times  \cal{B}_{\mu^{+}\mu^{-}}$ as a function of $m_{A^{0}}$. The shaded areas show the regions around the $J/\psi$ and the $\psi(2S)$ resonances excluded from the search.
  \label{Fig:3}
\end{figure}

\begin{figure}[htb]
  \centering

  \includegraphics[width=0.5\textwidth]{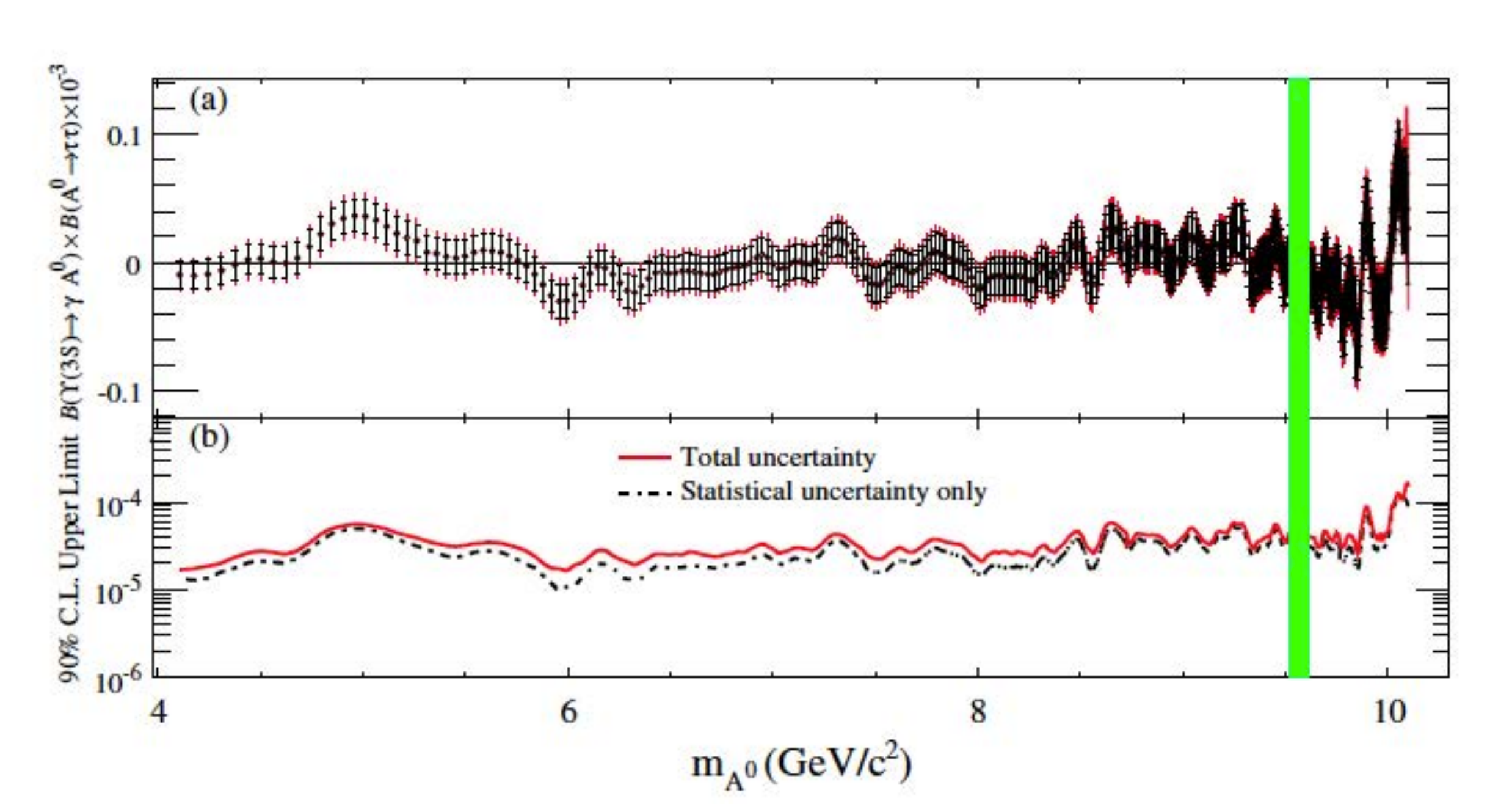}
   \caption{(a) Product branching fractions as a function of the Higgs mass. (b) The corresponding 90\% C.L. upper limits on the product of the branching fractions versus the Higgs mass values. }
  \label{Fig:4}
\end{figure}

\begin{figure}[htb]
  \centering

  \includegraphics[width=0.47\textwidth]{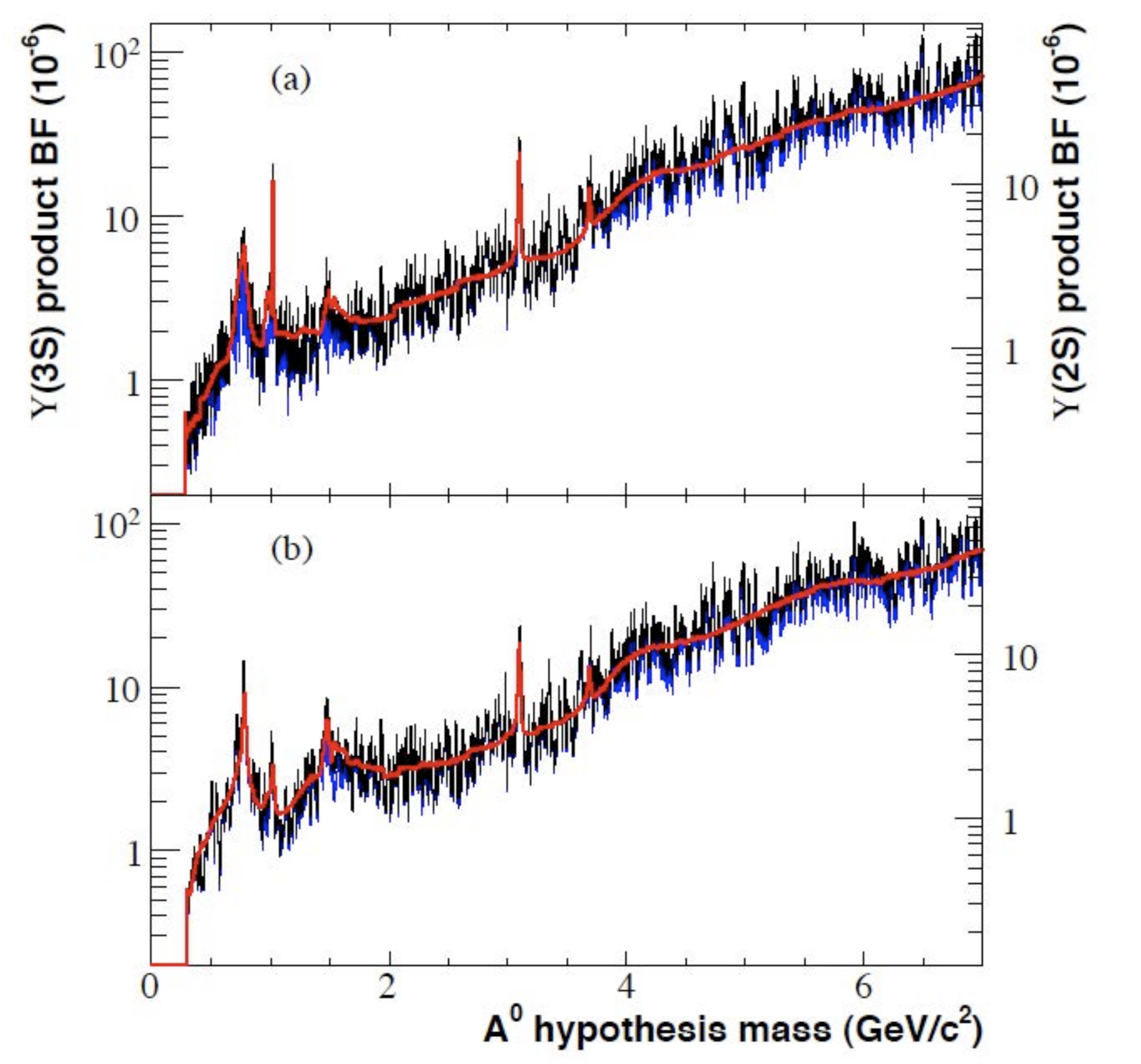}

 \caption{90 \% C.L. upper limits on product branching fractions (BF) (left axis) ${\cal B}  (\Upsilon(3S)\to \gamma A^{0})~{\cal B}(A^{0}\to hadrons)$ and (right axis) ${\cal B}  (\Upsilon(2S)\to \gamma A^{0})~{\cal B}(A^{0}\to hadrons)$ for (a) CP-all analysis and (b) CP-odd analysis.}
  \label{Fig:5}
\end{figure}

In recent years a number of theoretical models  \cite{Dermisek:2006py} 
predicted for the existence of a light CP-odd
Higgs boson $A^{0}$  related to the Next-to-Minimal Supersymmetric Model (NMSSM). Direct
searches constrain the mass of $m_{A^{0}} < 2m_{b}$, where $m_{b}$ is the b quark and the decay of $A^{0} \to b\bar{b}$
is forbidden. Of particular interest is to search the lightest CP-odd Higgs boson in $\Upsilon$
decays such as $\Upsilon(nS) \to \gamma A^{0}$, where $A^{0} \to SM $ particles. In these $\Upsilon(nS)$ transitions low mass Dark Matter Candidate ($\chi$) can be also directly produced (i.e. $\Upsilon(nS)\to \chi \chi$). The large data sets available at \babar experiment with more than 500 $fb^{-1}$
of data at  the $\Upsilon (4S), ~\Upsilon (3S), \Upsilon (2S)$ resonances and just below the  $ \Upsilon (4S)$ resonance
allow us to place stringent constraints on such theoretical models.\\
We searched for light Higgs in \babar using two-body-radiative decay on $\Upsilon$ states. The key experimental signature is the monochromatic
photon in the CM frame $E^{*}_{\gamma}=\frac{m_{\Upsilon}-m^{2}_{A^{0}}}{2m_{\Upsilon}}$. With this analysis technique we searched for  the
following transitions:
\begin{itemize}
\item $\Upsilon (2S,3S) \to \gamma A^{0},~A^{0}\to \mu^{+}\mu^{-}$  \cite{Aubert:2009cp} 
\item   $\Upsilon (2S,3S) \to \gamma A^{0},~A^{0}\to \tau^{+}\tau^{-}$\cite{Aubert:2009cka} 
\item    $\Upsilon (2S,3S) \to \gamma A^{0},~A^{0}\to \mathrm{hadrons}$\cite{Lees:2011wb} 
\end{itemize}
In all the searches we did not find any evidence of signals but upper limits versus hypothesis mass have been extracted the results are shown
in Fig.~\ref{Fig:3} for the $A^{0}\to \mu^{+}\mu^{-}$, in Fig. \ref{Fig:4} for the $A^{0}\to \tau^{+}\tau^{-}$ (where $\tau$ is reconstructed both in $\tau \to e \nu\bar{\nu}$ and  $\tau \to \mu \nu\bar{\nu}$) and in  Fig. \ref{Fig:5} for the  $A^{0}\to \mathrm{hadrons}$.

\section{Conclusion}
We have searched for evidence of dark sector candidates and evidence of CP-odd light Higgs in the $\Upsilon (2S) $ and $\Upsilon (3S) $
data sample at \babar. We did not observe any significant signal but more stringent limit have been set on space parameters of NP model.

% ****************************************************************************
% BIBLIOGRAPHY AREA
% ****************************************************************************

{\raggedright
\begin{footnotesize}
% IF YOU DO NOT USE BIBTEX, USE THE FOLLOWING SAMPLE SCHEME FOR THE REFERENCES
% ----------------------------------------------------------------------------

% ----------------------------------------------------------------------------

% IF YOU USE BIBTEX,
% - DELETE THE TEXT BETWEEN THE TWO ABOVE DASHED LINES
% - UNCOMMENT THE NEXT TWO LINES AND REPLACE 'smith_joe.bib' WITH YOUR
%   FILE(S)

% \bibliographystyle{DISproc}
% \bibliography{smith_joe.bib}
\end{footnotesize}
}

% ****************************************************************************
% END OF BIBLIOGRAPHY AREA
% ****************************************************************************

\end{document}